\documentclass[%
 aip,
 amsmath,amssymb,
 reprint,%
]{revtex4-1}

\usepackage{graphicx}
\usepackage{dcolumn}
\usepackage{bm}
\usepackage{subcaption}

\usepackage[utf8]{inputenc}
\usepackage[T1]{fontenc}
\usepackage{mathptmx}
\usepackage{amssymb}
\usepackage{etoolbox}
\usepackage{hyperref}
\usepackage{lineno}
\usepackage{multirow}
\usepackage{booktabs}

\begin{document}


\title[]{Revealing Secondary Particle Signatures \\in Muography Based on the Point of Closest Approach Algorithm}


\author{Rongfeng Zhang}

\affiliation{State Key Laboratory of Nuclear Physics and Technology, School of Physics, Peking University\\  Beijing, 100871, China 
\\This manuscript has been submitted to the Journal of Applied Physics for consideration.}

\author{Zibo Qin}

\affiliation{State Key Laboratory of Nuclear Physics and Technology, School of Physics, Peking University\\  Beijing, 100871, China 
\\This manuscript has been submitted to the Journal of Applied Physics for consideration.}

\author{Cheng-en Liu}

\affiliation{State Key Laboratory of Nuclear Physics and Technology, School of Physics, Peking University\\  Beijing, 100871, China
\\This manuscript has been submitted to the Journal of Applied Physics for consideration.}

\author{Qite Li}
\email[Contact author: ]{liqt@pku.edu.cn}
\affiliation{State Key Laboratory of Nuclear Physics and Technology, School of Physics, Peking University\\  Beijing, 100871, China 
\\This manuscript has been submitted to the Journal of Applied Physics for consideration.}

\author{Yong Ban}
\affiliation{State Key Laboratory of Nuclear Physics and Technology, School of Physics, Peking University\\  Beijing, 100871, China 
\\This manuscript has been submitted to the Journal of Applied Physics for consideration.}

\author{Chen Zhou}
\affiliation{State Key Laboratory of Nuclear Physics and Technology, School of Physics, Peking University\\  Beijing, 100871, China 
\\This manuscript has been submitted to the Journal of Applied Physics for consideration.}

\author{Qiang Li}
\email[Contact author: ]{qliphy0@pku.edu.cn}
\affiliation{State Key Laboratory of Nuclear Physics and Technology, School of Physics, Peking University\\  Beijing, 100871, China 
\\This manuscript has been submitted to the Journal of Applied Physics for consideration.}

\date{\today}

\begin{abstract}

This work reinterprets so-called ‘noise’ in cosmic ray imaging, indicating that the data of reconstructed Points of Closest Approach (PoCA points) outside the volume of interest defined by traditional tomography methods contain valuable physical information that has been traditionally disregarded. Through analysis of data from the detection system of four resistive plate chambers (RPCs) and Monte Carlo simulations employing energy deposition weighting for coordinate determination, we confirm that these points physically originate from the interaction between muons and the material above the detection system, particularly the roof, resulting in the production of secondary particles. The research yields two principal findings: first, in the four-layer compliance measurement system, the position recording of the first layer can be from secondary particles generated by cosmic rays, while the records from the three layers below represent the actual trajectories of cosmic rays; second, the roof structure significantly impacts the distribution of PoCA points at detector positions, where quantitative analysis demonstrates a strong correlation between roof thickness and the number of reconstructed PoCA points --- a relationship that can be precisely measured through $z$-coordinate distribution analysis in specific intervals. Due to the varying performances of different roofing materials in this analytical method, this approach holds significant potential for development into a new tomography technique.

\end{abstract}

\maketitle

\section{\label{sec1}INTRODUCTION}


Muon tomography was first proposed in 2003\cite{Borozdin_Hogan_Morris_Priedhorsky_Saunders_Schultz_Teasdale_2003}, with the PoCA algorithm representing one of the most widely used methods. The imaging principle simplifies mutiple Coulomb scattering of cosmic ray muons, assuming each muon's scattering occurs at a single point corresponding to the closest approach between its incoming and outgoing trajectories.\cite{Bonomi_Checchia_D’Errico_Pagano_Saracino_2020} Each PoCA point is weighted according to its deflection angle, where larger scattering angles receive higher weights due to their stronger correlation with high-density materials. Object images are obtained through pixel-based reconstruction.

In practical muon imaging applications, the naturally occurring cosmic ray flux (comprising both muons, electrons, and other components) is typically utilized. Recent advances in PoCA-based imaging research have primarily developed along two parallel tracks: (1) enhanced information extraction through innovative detector system designs (Ref.~\cite{He_Luo_Liu_Zou_Zhang_Xiao_Huang_Wang_2024,Chen_Liu_Wang_Han_Gong_2025}), and (2) improved data processing via theoretical refinements (Ref.~\cite{Wang_Fan_Wang_Wen_2025,Abhishek_Karnam_Kashyap_Mohanty_2025}) or machine learning approaches (Ref.~\cite{Chen_Liu_Wang_Han_Gong_2025,Vinodkumar_Avots_Ozcinar_Anbarjafari_2025,López,Luo_Zhang_Yin_Zeng_Feng_Feng_Du_Xiao_Wang_2025}). Among them, imaging analyses have focused exclusively on the sample regions (known as volume of interest) while treating points reconstructed in other spatial domains as noise. This work demonstrates that the PoCA points reconstructed within the detector contain physically meaningful signatures, providing an underutilized source of information for tomography.


\section{\label{sec2}EXPERIMENTAL SETUP}

\begin{figure*}[htbp]
  \centering
  \begin{subfigure}[b]{0.64\textwidth}
    \includegraphics[width=\linewidth]{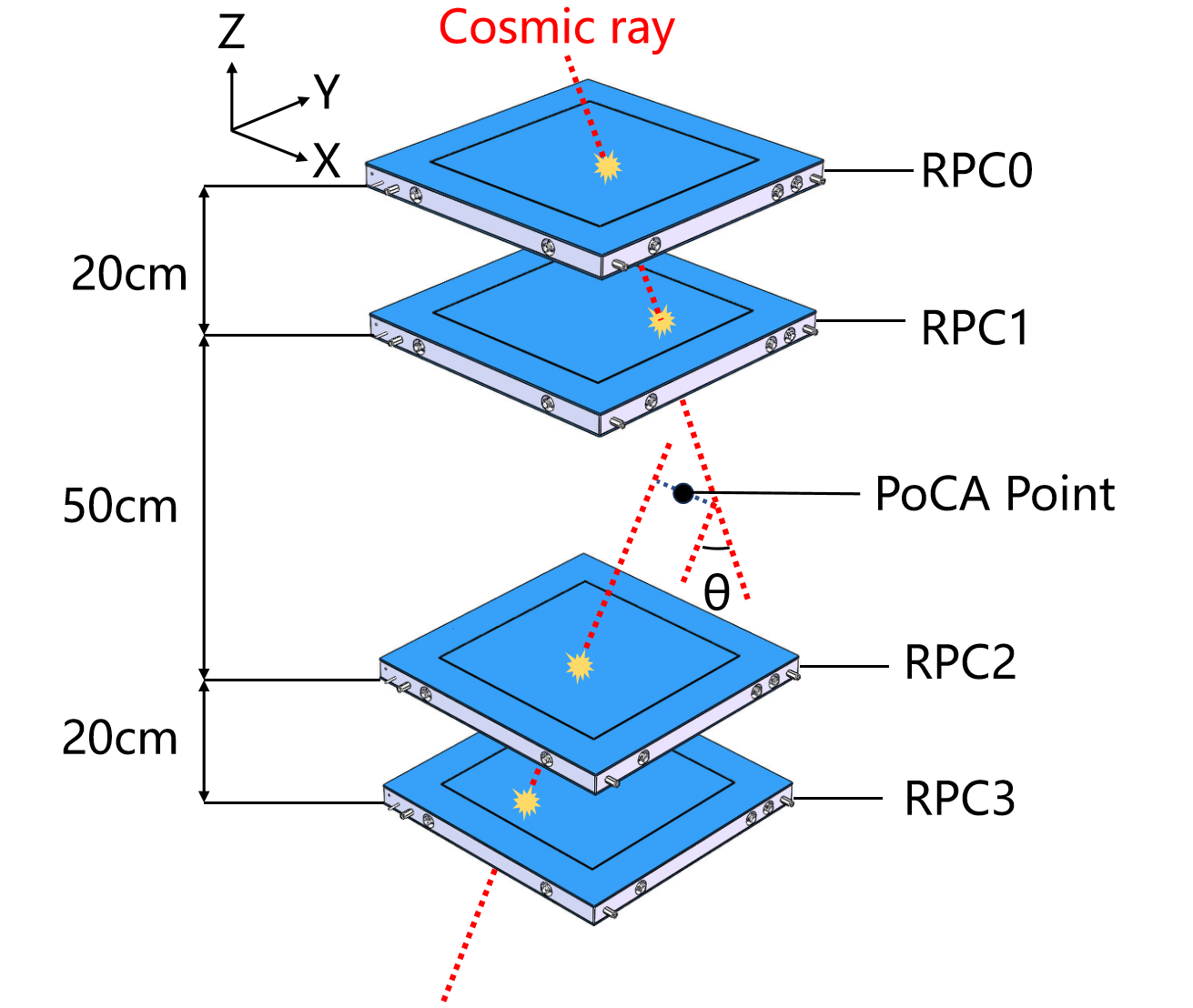}
    \caption{}\label{fig:1a}
  \end{subfigure}
  \hfill
  \begin{subfigure}[b]{0.32\textwidth}
    \includegraphics[width=\linewidth]{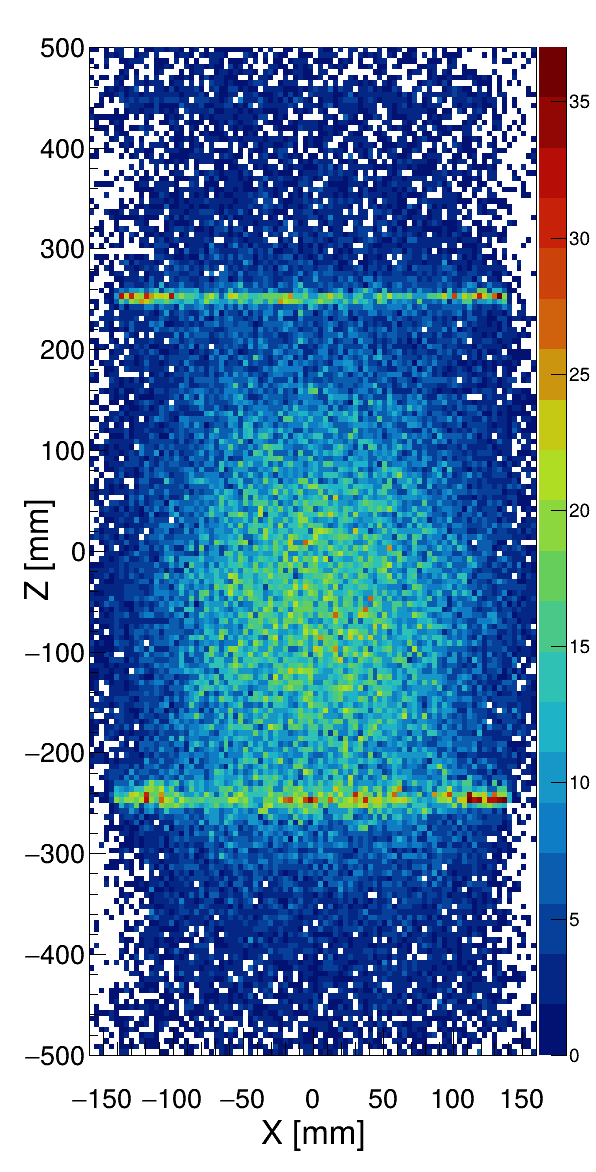}
    \caption{}\label{fig:1b}
  \end{subfigure}
  \caption{(a) Main body of the detection system. Here, ancillary components such as the high-voltage power supply, ventilation device, and electronic plugins are omitted. (b) PoCA point distribution in the X-Z direction. The color scale corresponds to the number of PoCA points recorded per pixel, with longer-wavelength colors indicating higher values.}
  \label{fig:both}
\end{figure*}

This study originates from the PKMu collaboration’s program~\cite{Yu2024, doi:10.1142/S0217732325300083} to search for muon–dark matter scattering, which necessitates precise identification and simulation of various scattering signatures in cosmic-ray imaging system. The cosmic-ray imaging system comprises four resistive plate chambers (RPCs) with 28 × 28 cm$^2$ active area arranged with inter layer spacings of 20\,cm, 50\,cm, and 20\,cm between the bottom surfaces (Fig.~\ref{fig:1a}). 


Each RPC provides particle position through the LC delay-line readout method.~\cite{Li_Ye_Ji_Wen_Liu_Ge_2013, LI201222} When a particle triggers an ionization avalanche process within the detector, it will generate an induced signal on nearby readout strips. The readout strips are connected to LC delay lines that provide a fixed time delay to the signals. The time difference between signal arrivals at both ends of the line determines the incident position. Influenced by the electronics noise and electromagnetic interference, the resolution of each RPC is approximately 0.7\,mm. There is also a signal output without delay in each RPC, known as the T channel, which is used for triggering. This study considers only events with simultaneous signals and position recoded in all four RPCs, a requirement of the PoCA algorithm.

1,1,1,2-Tetrafluoroethane (also known as R-134a) was selected as the working gas for the RPC. During the experiment, the set of voltage for RPCs is to ensure that the two-dimensional position reconstruction efficiency  for each RPC is approximately 85\%. After 63 days of targetless measurements, approximately 1.18 million events were accumulated for analysis.


\section{\label{sec3}SIMULATION SETUP}


The GEANT4 11.1.2 toolkit\cite{Agostinelli2003,Allison2006,Allison2016} performed Monte Carlo simulations, while the PKMu collaboration developed the detector geometry and maintains its open-source implementation on GitHub.\cite{PKMUON2024Config} Each RPC occupies a cuboid volume of 300$\times$300$\times$30\,mm$^3$ and sets a sensitive area that captures particle type, deposited energy, and spatial coordinates during particle interactions. The horizontal position coordinates are obtained by calculating the energy deposition weighted average of each particle's horizontal coordinate. The $z$-positions are fixed at 450\,mm, 250\,mm, $-$250\,mm, and $-$450\,mm. The simulated $x$ and $y$ coordinates are modified by applying random Gaussian offsets based on the experimentally determined detector resolution.


\begin{figure}[htbp]
  
    \centering
    \includegraphics[width=\linewidth]{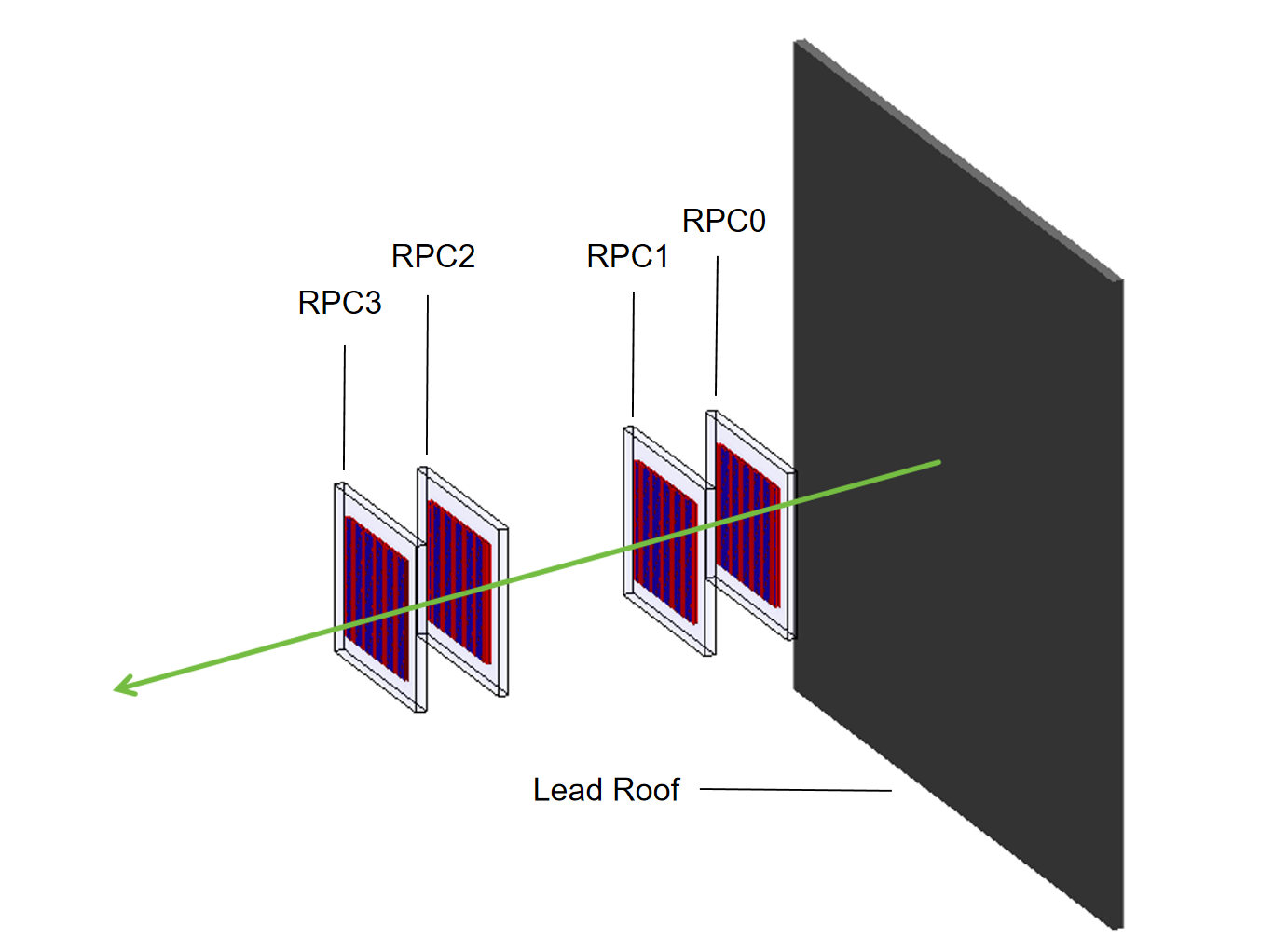}
    \caption{\label{fig:2}GEANT4 simulation setup: Roof (gray) and detector (white wireframe). The green arrow represents the direction of incidence of the particles.}

\end{figure}


The cosmic-ray showers were generated using the CRY\cite{Hagmann_Lange_Wright_2007} software package, which reproduces the characteristic energy spectrum and angular distributions observed in the Beijing region at sea level. The initial simulation (Run 0) configuration uses a plane particle generator of 300$\times$300\,mm$^2$ placed directly at the top of RPC0. The simulation utilizes the FTFP\_BERT physics list, which is a pre-packaged physics list provided by Geant4. It includes hadronic processes, decay processes, and other standard processes. For electromagnetic processes, we have replaced the original constructor with G4EmStandardPhysics\_option4. Additionally, G4StepLimiterPhysics has been added to impose limits on particle step lengths. These physical processes essentially cover the interactions that may occur between cosmic rays and detection systems. 

By dividing the pixels, we calculated the positions of the PoCA points using experimental data and illustrated their distribution in the X-Z direction (Fig.~\ref{fig:1b}). To explain the phenomenon of clustered PoCA points around the detector in the experiment, the simulation combined roofs of varying thicknesses (Run 1). Since the actual indoor environment is shielded by multiple layers of roofing, most of the low-energy secondary particles should be absorbed. For the purpose of simplifying the model for preliminary research attempts, we referenced the spacing settings of the two middle detectors, maintaining a constant distance of 500\,mm between the top of the roof and the top of RPC0. The dimensions of the roof were matched to the generator area, while the particle plane was enlarged to 1515$\times$1515\,mm$^2$ to ensure unbiased sampling of the incident angles (Fig.~\ref{fig:2}). In other words, particles generated from the edges of the newly formed plane can pass through the upper two layers of the RPC at the same incident angle as those generated from the edges of the previously formed plane. 

In Run 0, we generated approximately 1.24 million event data points, whereas for each roofing configuration in Run 1, we produced about one-tenth of the data of Run 0. Both experiment and simulation data were saved in ROOT format\cite{Brun_Rademakers_1997} for analysis.


\section{\label{sec4}DATA ANALYSIS}

\begin{figure}[htbp]
    \includegraphics[width=\linewidth]{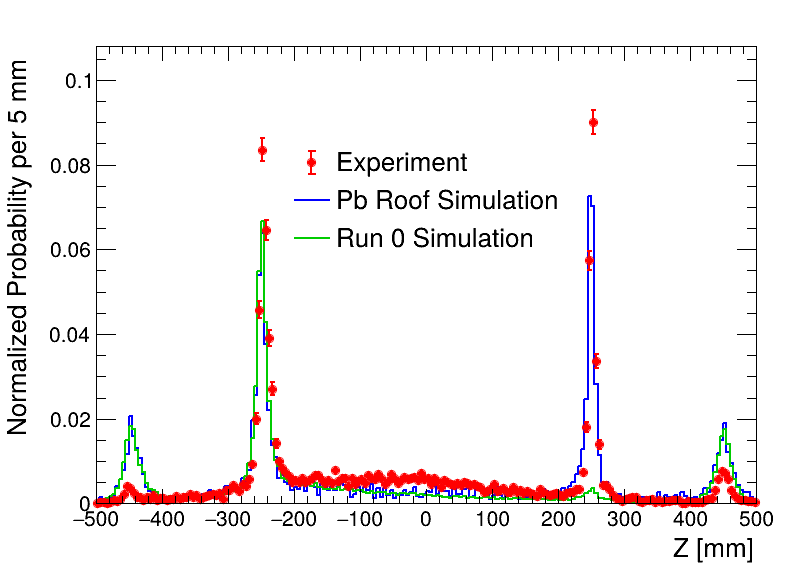}
    \caption{Comparison of the filtered PoCA $z$-distributions ($\theta > 0.2$\,rad, $\Delta d < 10$\,mm, within $300\times300\times1000$\,mm$^3$ volume) from the experiment and simulations. The values in the vertical axis labels represent the interval width of each bin on the horizontal axis. The experimental data is represented by points with error bars. The data for the lead roof shown in the image comes from simulations of a 10\,mm thick lead roof.}
    \label{fig:3}
\end{figure}

\subsection{\label{sec4:A}Simulation-experiment discrepancy}


Although RPCs are made from lightweight materials such as aluminum and glass, they still possess a density greater than that of air, resulting in a larger average scattering angle of cosmic rays. Compared to the multiple scattering processes occurring in air, the processes within the detector are more likely to achieve larger scattering angles with fewer scattering occurrences, which is manifested by the small spatial separation $\Delta d$ between the incident and outgoing trajectories. To effectively isolate these signals from the cosmic-ray multiple scattering background in air, the analysis employed stringent selection criteria: $\theta > 0.2$\,rad, $\Delta d < 10$\,mm, within $300\times300\times1000$\,mm$^3$ volume. Among them, $\theta$ represents the scattering angle calculated using the PoCA algorithm, and the selection of the size in the $z$-direction is considered for the convenience of bin segmentation. A special normalization method was employed during the plotting process in order to prevent the selection criteria from introducing additional discrepancies. For the experimental data, the normalization base is set to one percent of the total number of PoCA points (which is close to the number of filtered PoCA points). For the simulated data, the normalization base is chosen as the total number of PoCA points multiplied by a conversion factor that is a constant value. The calculation of this conversion factor is based on the simulated data of a 10\,mm thick lead roof, ensuring that after normalization, the proportion of the filtered PoCA points in this set matches that of the experimental data. This can be regarded as a correction after considering the experimental detection efficiency. The existence of the two-dimensional position reconstruction efficiency may be the reason for the differing proportions of events after the experimental and simulation selection. The method for calculating the normalized probability in the image involves dividing the number of events in each bin, after filtering, by a specially processed denominator.

As shown in Fig.~\ref{fig:3}, while the experimental data showed pronounced clustering within the interval [235, 265]\,mm, also known as RPC1's position, this feature was notably absent in the Run 0 simulation result. The situation within the interval [\-465, \-435]\,mm (also referred to as the RPC0 region) appears to offer a hint. Ordinary scattering would not account for this phenomenon, as such events would be recorded as a straight line passing through the detection system (assuming no trajectory deviations occur from subsequent cosmic rays). However, if this event is interpreted as a positional recording deviation from RPC1, it becomes straightforward to obtain this PoCA reconstruction result. Through the examination of the simulated data, it can be found that in a single event, a single RPC may record multiple particles, which is likely a reason for the deviation in the RPC position recording. Therefore, if the distribution of the PoCA point at the RPC1 position is understood as cosmic rays passing straight through the detection system, but there are issues with the recordings at the RPC0 position, then the Run 0 simulation lacks the secondary particle source, which may be the roof, compared to the experiment.


\subsection{\label{sec4:B}Roof simulations}

\begin{figure}[htbp]
    \includegraphics[width=\linewidth]{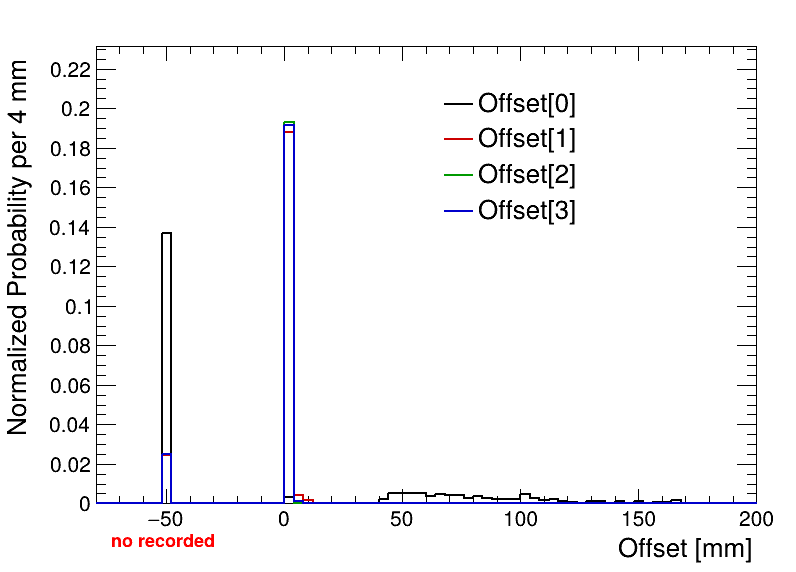}
    \caption{The offset distance between the actual recorded positions of RPCs and the primary particle impact positions on the RPCs. The offset labels correspond to the RPC labels. Negative values indicate that the initial particle did not impact the RPC. The vertical axis markings are the same as in Fig.~\ref{fig:3}.}
    \label{fig:4}
\end{figure}
\begin{figure}[htbp]
    \includegraphics[width=\linewidth]{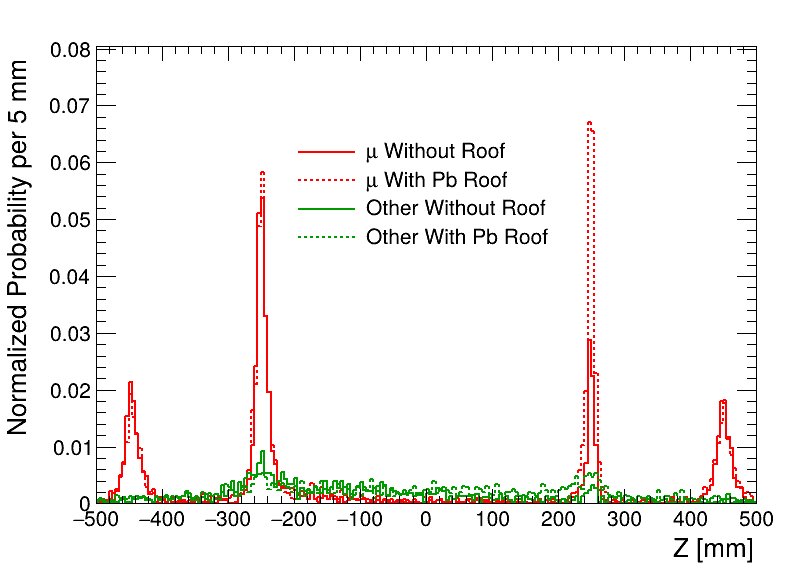}
    \caption{Comparison of the filtered PoCA $z$-distributions ($\theta > 0.2$\,rad, $\Delta d < 10$\,mm, within $300\times300\times1000$\,mm$^3$ volume) from different primary particles.The term 'Without Roof' means that the position where the roof was originally placed in the Run 1 simulation has been filled with air. The lead roof is also referenced at 10\,mm. The axis markings are the same as in Fig.~\ref{fig:3}.}
    \label{fig:5}
\end{figure}

Fig.~\ref{fig:3} also illustrates the simulation data for a 10\,mm lead roof, indicating that the lead roof clearly alters the PoCA $z$-distribution in the RPC1 region. In this simulation, the trajectories of the primary particles were tracked, and the offset between the hit position of the primary particles and the recorded actual position was calculated, as shown in Fig.~\ref{fig:4}. For cases where the initial particles were not recorded by the detector, the offset value was fixed at $-$50\,mm for clearer differentiation. In order to focus on the PoCA points at the RPC1 location, the plotting only utilized events with the $z$-coordinate of the PoCA points within the range of [235, 265]\,mm in Fig.~\ref{fig:3}, based on the original selection criteria. The results indicate that the manner in which secondary particles influence the PoCA point slightly deviates from previous inferences. In the filtered events, the vast majority of primary particles were found to pass only through the three lower RPC layers, while the records from RPC0 were produced by mainly electrons resulting from interactions between primary particles and the roof. The particle generation plane in the Run 0 simulation has geometrically ruled out this possibility; therefore, a control simulation filled solely with air was set up in the simplified roof simulations. Fig.~\ref{fig:5} categorizes the filtered events based on the types of primary particles as simulated. The distribution of the three types of particles within each simulation shares a common normalization constant, as mentioned in the previous section. The results indicate that the PoCA points reconstructed in RPC1 are primarily caused by muon events, and the roof will further increase the proportion of PoCA points reconstructed on RPC1.

\begin{figure}
    \includegraphics[width=1.\columnwidth]{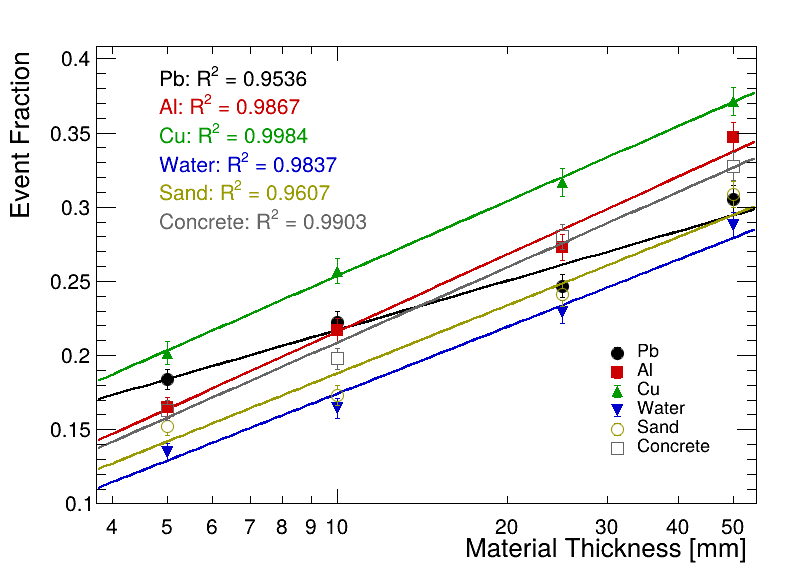}
    \caption{The correlation between the thickness of roofs and the corresponding event fractions. The error bars represent the statistical uncertainty arising from fluctuations in event counts. The horizontal axis is displayed on a logarithmic scale to emphasize the variations across different thicknesses. }
    \label{fig:6}
\end{figure}

To further investigate this phenomenon, simulations with varying roof thicknesses were conducted. The arrangement of roofs of different thicknesses is referenced in Section \ref{sec3}. Water and concrete utilized the material definitions provided by Geant4, while sand was replaced with silica included in Geant4. For quantitative analysis of the PoCA point clustering distribution at RPC1's position, the integral of the PoCA $z$-distribution within the [235, 265] \,mm interval was introduced as an effective metric. This event fraction from the RPC1 region is proportional to the number of PoCA points reconstructed within RPC1 after screening in Section \ref{sec4:A}. Fig.~\ref{fig:6} demonstrates a remarkably strong positive correlation between the event fraction and the thickness of roofs. Except for lead, the thickness and event fractions of various other materials show a very similar relationship. Moreover, unlike traditional imaging methods, the positional relationship of the several fitted lines in Fig.~\ref{fig:6} does not exhibit a very clear correlation with the atomic number and surface density of the materials. The special performance of the lead roof is currently speculated to be influenced by the absorption effect of lead. The complete dataset is recorded in Table \ref{tab:table1}.

According to Ref.\cite{Galgóczi_2020}, differences exist in the secondary particles generated when muons interact with different materials. Although only the emission angles of secondary particles from a lead target were explicitly provided, these data still suggest that distinctions in secondary particle production occur across materials, thereby offering support for the present study.


\begin{table*}
\caption{\label{tab:table1}Roof thickness versus event fractions (proportional to the number of PoCA points reconstructed within RPC1 after screening in Section \ref{sec4:A}) from simulations. All data was obtained with consistent event selection criteria.}
\begin{ruledtabular}
\begin{tabular}{c ccc ccc}
\multirow{2}{*}{Thickness (mm)} & \multicolumn{3}{c}{Metals' Event Fraction} & \multicolumn{3}{c}{Non-metals' Event Fraction} \\ 
\cmidrule(lr){2-4} \cmidrule(lr){5-7}
 & Pb & Cu & Al & Water & Sand & Concrete \\ \hline
50 & 0.305$\pm$0.009 & 0.371$\pm$0.010 & 0.347$\pm$0.010 & 0.287$\pm$0.009 & 0.309$\pm$0.009 & 0.327$\pm$0.009 \\
25 & 0.247$\pm$0.008 & 0.317$\pm$0.009 & 0.273$\pm$0.009 & 0.229$\pm$0.008 & 0.241$\pm$0.008 & 0.280$\pm$0.009 \\
10 & 0.225$\pm$0.007 & 0.257$\pm$0.009 & 0.217$\pm$0.008 & 0.164$\pm$0.007 & 0.173$\pm$0.007 & 0.198$\pm$0.007 \\
5  & 0.184$\pm$0.007 & 0.202$\pm$0.008 & 0.165$\pm$0.007 & 0.135$\pm$0.006 & 0.152$\pm$0.006 & 0.162$\pm$0.007 \\
\end{tabular}
\end{ruledtabular}
\end{table*}

\section{\label{sec5}SUMMARY}

While conventional cosmic-ray scattering imaging methods typically focus on scattering occurring exclusively within the region of interest (ROI) between incident and exit detectors, this work demonstrates that materials located above the imaging system can also significantly influence the reconstructed scattering point. After a special selection (see section ~\ref{sec4:A}), the number of PoCA points reconstructed on RPC1 relates to certain physical processes that occur before the muons enter the detection system. The analysis shows that these reconstructed points primarily reflect a specific class of processes --- where the hit locations of muons are recorded by the bottom three layers of the detector, while the position recorded by the first layer is caused by secondary particles resulting from interactions between the muons and the matter in front of the detection system. We redefine the proportion of the number of particles located at RPC1 after normalization as the event fraction and use this as a metric to measure the aforementioned phenomenon in subsequent simulations of roofs made of different materials and thicknesses. The results indicate that this metric has a strong positive correlation with the thickness of the roof and is capable of distinguishing between different materials. This study broadens the application methods of the PoCA algorithm and has the potential to develop into a new imaging technique. In subsequent research, we will advance the quantitative explanation of this phenomenon and explore its association with certain physical properties of materials. These investigations are built upon the capability of this method to reveal the influence of secondary particles, enabling material sensitivity above cosmic-ray imaging detectors --- unlocking potential applications in mineral exploration, archaeological survey, and civil engineering, with further performance gains anticipated from artificial muon sources.


\section{ACKNOWLEDGEMENTS}

This work is supported in part by the National Natural Science Foundation of China under Grants No. 12325504, and No. 12061141002.

\section{DATA AVAILABILITY}
All data included in this study are available upon request by contact with the corresponding author.

\nocite{*}
\bibliography{aipsamp}

\end{document}